\documentclass[sigconf]{acmart}

\AtBeginDocument{%
  \providecommand\BibTeX{{%
    \normalfont B\kern-0.5em{\scshape i\kern-0.25em b}\kern-0.8em\TeX}}}

\copyrightyear{2020}
\acmYear{2020}
\setcopyright{iw3c2w3}
\acmConference[WWW '20]{Proceedings of The Web Conference 2020}{April 20--24, 2020}{Taipei, Taiwan}
\acmBooktitle{Proceedings of The Web Conference 2020 (WWW '20), April 20--24, 2020, Taipei, Taiwan}
\acmPrice{}
\acmDOI{10.1145/3366423.3380239}
\acmISBN{978-1-4503-7023-3/20/04}

\usepackage{amsmath,amssymb,amsfonts}
\usepackage{algorithmic}
\usepackage{graphicx}
\usepackage{textcomp}
\usepackage{xcolor}
\usepackage{enumitem}
\usepackage[font={small}]{caption}
\usepackage{wrapfig}
\usepackage[T1]{fontenc}
\usepackage{mathtools}
\usepackage{enumitem,comment,wasysym}
\usepackage{color, soul}
\usepackage{pbox}
\usepackage{listings}
\usepackage{tikz,subfigure}
\usepackage{url}
\usepackage{booktabs}
\usepackage{multirow}
\usepackage{xspace}

\usepackage{breakurl}
\usepackage{hyperref}
\usepackage{cleveref}

\usepackage[subtle]{savetrees}

\definecolor{dkgreen}{rgb}{0,0.6,0}
\definecolor{gray}{rgb}{0.5,0.5,0.5}
\definecolor{mauve}{rgb}{0.58,0,0.82}
\definecolor{editorGray}{rgb}{0.95, 0.95, 0.95}
\definecolor{editorOcher}{rgb}{1, 0.5, 0}
\definecolor{editorGreen}{rgb}{0,0.6,0}
\definecolor{darkgreen}{RGB}{0,90,90}
\definecolor{lightgray}{rgb}{0.95, 0.95, 0.95}
\definecolor{darkgray}{rgb}{0.4, 0.4, 0.4}
\definecolor{orange}{rgb}{1,0.45,0.13}		
\definecolor{olive}{rgb}{0.17,0.59,0.20}
\definecolor{brown}{rgb}{0.69,0.31,0.31}
\definecolor{purple}{rgb}{0.38,0.18,0.81}
\definecolor{lightblue}{rgb}{0.1,0.57,0.7}
\definecolor{lightred}{rgb}{1,0.4,0.5}
\definecolor{dkgreen}{rgb}{0,0.6,0}
\definecolor{gray}{rgb}{0.5,0.5,0.5}
\definecolor{mauve}{rgb}{0.58,0,0.82}
\definecolor{editorGray}{rgb}{0.95, 0.95, 0.95}
\definecolor{editorOcher}{rgb}{1, 0.5, 0}
\definecolor{editorGreen}{rgb}{0,0.6,0}

\definecolor{MyMaroon}{HTML}{D50000}
\definecolor{MyYellow}{HTML}{FFD600}
\definecolor{MyGreen}{HTML}{00C853}

\definecolor{MyTeal}{HTML}{009688}
\definecolor{MyOrange}{HTML}{FF5722}
\definecolor{MyGray}{HTML}{E0E0E0}
\definecolor{MyIndigo}{HTML}{3F51B5}
\definecolor{MyPurple}{HTML}{9C27B0}
\definecolor{MyBlack}{HTML}{000000}
\definecolor{MyBlue}{HTML}{3F51B5}
\newcommand{\JS}{JavaScript\xspace}
\newcommand{\etc}{etc.}
\newcommand{\eg}{e.g. }
\newcommand{\ie}{i.e. }

\newcommand{\pg}{PageGraph\xspace}
\newcommand{\point}[1]{\par\smallskip\noindent\textbf{#1.}}

\newcommand{\Accuracy}{\hybridCVAccuracy{}\xspace}

\newcommand{\NumTotalNewResources}{3,349\xspace}
\newcommand{\NumTotalUniqueResources}{1,771\xspace}

\newcommand{\NumTotalUniqueResourcesAL}{511\xspace}

\newcommand{\NumTotalUniqueResourcesHU}{586\xspace}

\newcommand{\NumTotalUniqueResourcesLK}{674\xspace}

\newcommand{\NumDomainsCrawledLKTotal}{890\xspace}

\newcommand{\NumDomainsCrawledHUTotal}{935\xspace}

\newcommand{\NumDomainsCrawledALTotal}{740\xspace}

\newcommand{\NumDomainsCrawledAllTotal}{2,565\xspace}

\newcommand{\NumPagesCrawledLKUnique}{2,196\xspace}

\newcommand{\NumPagesCrawledHUUnique}{2,287\xspace}

\newcommand{\NumPagesCrawledALUnique}{1,805\xspace}

\newcommand{\NumPagesCrawledAllUnique}{6,288\xspace}
\newcommand{\NumPagesCrawledAllTotal}{6,475\xspace}

\newcommand{\AllImagesLKUnique}{47,092\xspace}

\newcommand{\AllImagesHUUnique}{46,687\xspace}

\newcommand{\AllImagesALUnique}{38,763\xspace}

\newcommand{\AllImagesAllUnique}{132,542\xspace}

\newcommand{\ImagesDirectFromListsLKUnique}{1,451\xspace}
\newcommand{\ImagesDirectFromListsLKUniquePct}{3.1\%\xspace}

\newcommand{\ImagesDirectFromListsHUUnique}{1,479\xspace}
\newcommand{\ImagesDirectFromListsHUUniquePct}{3.2\%\xspace}

\newcommand{\ImagesDirectFromListsALUnique}{2,553\xspace}
\newcommand{\ImagesDirectFromListsALUniquePct}{6.6\%\xspace}

\newcommand{\ImagesDirectFromListsCombinedUnique}{5,483\xspace}
\newcommand{\ImagesDirectFromListsCombinedUniquePct}{4.1\%\xspace}

\newcommand{\FramesDirectFromListsLKUnique}{296\xspace}
\newcommand{\FramesDirectFromListsLKUniquePct}{35.1\%\xspace}

\newcommand{\FramesDirectFromListsHUUnique}{209\xspace}
\newcommand{\FramesDirectFromListsHUUniquePct}{15.9\%\xspace}

\newcommand{\FramesDirectFromListsALUnique}{164\xspace}
\newcommand{\FramesDirectFromListsALUniquePct}{28.4\%\xspace}

\newcommand{\FramesDirectFromListsCombinedUnique}{669\xspace}
\newcommand{\FramesDirectFromListsCombinedUniquePct}{24.4\%\xspace}

\newcommand{\CombinedDirectFromListsLKUnique}{1,872\xspace}

\newcommand{\CombinedDirectFromListsHUUnique}{1,819\xspace}

\newcommand{\CombinedDirectFromListsALUnique}{2,850\xspace}

\newcommand{\CombinedDirectFromListsAllUnique}{6,541\xspace}

\newcommand{\BlockedMoreTotalPercent}{30.1\%\xspace}
\newcommand{\BlockedMoreUniquePercent}{27.1\%\xspace}

\newcommand{\BlockedMoreUniquePercentAL}{18\%\xspace}

\newcommand{\BlockedMoreUniquePercentHU}{32.2\%\xspace}

\newcommand{\BlockedMoreUniquePercentLK}{36\%\xspace}

\newcommand{\filterlistrulesAlbania}{201\xspace}
\newcommand{\filterlistrulesHungary}{1,407\xspace}
\newcommand{\filterlistrulesSriLanka}{69\xspace}
\newcommand{\filterlistNetworkRulesAlbania}{118\xspace}
\newcommand{\filterlistNetworkRulesHungary}{662\xspace}
\newcommand{\filterlistNetworkRulesSriLanka}{42\xspace}
\newcommand{\filterlistUpdatedAlbania}{8\xspace}
\newcommand{\filterlistUpdatedHungary}{3\xspace}
\newcommand{\filterlistUpdatedSriLanka}{22\xspace}

\newcommand{\filterlistCommitsAL}{3\xspace}
\newcommand{\filterlistStartMonthAL}{02-2019\xspace}
\newcommand{\filterlistCommitsPerMonthAL}{0.27\xspace}
\newcommand{\filterlistCommitsHU}{542\xspace}
\newcommand{\filterlistStartMonthHU}{12-2014\xspace}
\newcommand{\filterlistCommitsPerMonthHU}{8.9\xspace}
\newcommand{\filterlistCommitsLK}{16\xspace}
\newcommand{\filterlistStartMonthLK}{03-2016\xspace}
\newcommand{\filterlistCommitsPerMonthLK}{0.35\xspace}
\newcommand{\filterlistCommitsHI}{1,637\xspace}
\newcommand{\filterlistStartMonthHI}{05-2018\xspace}
\newcommand{\filterlistCommitsPerMonthHI}{81.85\xspace}
\newcommand{\filterlistCommitsGE}{11,982\xspace}
\newcommand{\filterlistStartMonthGE}{01-2014\xspace}
\newcommand{\filterlistCommitsPerMonthGE}{166.4\xspace}
\newcommand{\filterlistCommitsJA}{1,687\xspace}
\newcommand{\filterlistStartMonthJA}{05-2014\xspace}
\newcommand{\filterlistCommitsPerMonthJA}{24.8\xspace}

\newcommand{\perceptualCVPrecision}{95.5\,\%\xspace}
\newcommand{\perceptualCVRecall}{96.4\,\%\xspace}
\newcommand{\perceptualCVAccuracy}{95.9\,\%\xspace}

\newcommand{\perceptualOtherCrawlPrecision}{48.8\,\%\xspace}
\newcommand{\perceptualOtherCrawlRecall}{87.4\,\%\xspace}
\newcommand{\perceptualOtherCrawlAccuracy}{77.0\,\%\xspace}

\newcommand{\hybridCVPrecision}{92\,\%\xspace}
\newcommand{\hybridCVRecall}{75\,\%\xspace}
\newcommand{\hybridCVAccuracy}{97.6\,\%\xspace}

\newcommand{\NewALFilterlistRules}{387\xspace}
\newcommand{\NewHUFilterlistRules}{551\xspace}
\newcommand{\NewLKFilterlistRules}{372\xspace}
\newcommand{\NewTotalFilterlistRules}{1310\xspace}

\newcommand{\crawlOneTotalImages}{7,685\xspace}
\newcommand{\crawlOneImagesAdsRatio}{48\,\%\xspace}
\newcommand{\crawlOneTotalFrames}{465\xspace}
\newcommand{\crawlOneFramesAdsRatio}{77\,\%\xspace}

\newcommand{\crawlTwoTotalImages}{2,610\xspace}
\newcommand{\crawlTwoImagesAdsRatio}{14\,\%\xspace}
\newcommand{\crawlTwoTotalFrames}{1,034\xspace}
\newcommand{\crawlTwoFramesAdsRatio}{41\,\%\xspace}

\newcommand{\NumFramesObservedUniqueAL}{578\xspace}
\newcommand{\NumFramesObservedUniqueHU}{1,318\xspace}
\newcommand{\NumFramesObservedUniqueLK}{844\xspace}
\newcommand{\NumFramesObservedUniqueCombined}{2,740\xspace}

\newcommand{\ImagesDirectFromOnlyUsALUnique}{440\xspace}
\newcommand{\ImagesDirectFromOnlyUsALUniquePct}{17.2\%\xspace}
\newcommand{\ImagesDirectFromOnlyUsHUUnique}{547\xspace}
\newcommand{\ImagesDirectFromOnlyUsHUUniquePct}{37\%\xspace}
\newcommand{\ImagesDirectFromOnlyUsLKUnique}{510\xspace}
\newcommand{\ImagesDirectFromOnlyUsLKUniquePct}{35.1\%\xspace}

\newcommand{\ImagesDirectFromOnlyUsCombinedUnique}{1,497\xspace}
\newcommand{\ImagesDirectFromOnlyUsCombinedUniquePct}{27.3\%\xspace}

\newcommand{\FramesDirectFromOnlyUsALUnique}{20\xspace}
\newcommand{\FramesDirectFromOnlyUsALUniquePct}{12.2\%\xspace}
\newcommand{\FramesDirectFromOnlyUsHUUnique}{10\xspace}
\newcommand{\FramesDirectFromOnlyUsHUUniquePct}{4.8\%\xspace}
\newcommand{\FramesDirectFromOnlyUsLKUnique}{17\xspace}
\newcommand{\FramesDirectFromOnlyUsLKUniquePct}{5.7\%\xspace}

\newcommand{\FramesDirectFromOnlyUsCombinedUnique}{47\xspace}
\newcommand{\FramesDirectFromOnlyUsCombinedUniquePct}{7\%\xspace}

\newcommand{\CombinedDirectFromOnlyUsALUnique}{460\xspace}
\newcommand{\CombinedDirectFromOnlyUsHUUnique}{557\xspace}
\newcommand{\CombinedDirectFromOnlyUsLKUnique}{527\xspace}
\newcommand{\CombinedDirectFromOnlyUsCombinedUnique}{1,544\xspace}

\begin{document}
\sloppy

\title{Filter List Generation for Underserved Regions}

\author{Alexander Sj\"osten}
\affiliation{
  \institution{Chalmers University of Technology}
}

\author{Peter Snyder}
\affiliation{
  \institution{Brave Software}
}

\author{Antonio Pastor}
\affiliation{
  \institution{Universidad Carlos III de Madrid}
}

\author{Panagiotis Papadopoulos}
\affiliation{
  \institution{Brave Software}
}

\author{Benjamin Livshits}
\affiliation{
  \institution{Brave Software}
  \institution{Imperial College London}
}

\renewcommand{\shortauthors}{Sj\"osten and Snyder, et al.}

\setcounter{topnumber}{4}
\setcounter{bottomnumber}{4}
\setcounter{totalnumber}{4}

\begin{abstract}
  Filter lists play a large and growing role in protecting and assisting web users. The vast majority of popular filter lists are crowd-sourced, where a
  large number of people manually label resources related to undesirable   web resources (\eg ads, trackers, paywall libraries), so that they can be blocked by browsers and extensions.

  Because only a small percentage of web users participate in the generation   of filter lists, a crowd-sourcing strategy works well for blocking either
  uncommon resources that appear on ``popular'' websites, or resources that   appear on a large number of ``unpopular'' websites. A crowd-sourcing
  strategy will perform poorly for parts of the web with small ``crowds'', such as regions of the web serving languages with (relatively) few speakers.

  This work addresses this problem through the combination of two novel techniques: (i) deep browser instrumentation that allows for
  the accurate generation of request chains, in a way that is robust in
  situations that confuse existing measurement techniques, and (ii) an ad classifier that uniquely combines perceptual and page-context
  features to remain accurate across multiple languages.

  We apply our unique two-step filter list generation pipeline to three regions of the web that currently have poorly maintained
  filter lists: Sri Lanka, Hungary, and Albania. We generate new filter lists that complement existing filter lists. Our complementary lists
  block an additional~\NumTotalNewResources{} of ad and ad-related resources
  (\NumTotalUniqueResources{} unique) when applied to~\NumPagesCrawledAllTotal
  pages targeting these three regions.

  We hope that this work can be part of an increased effort at ensuring that the
  security, privacy, and performance benefits of web resource blocking can be
  shared with all users, and not only those in dominant linguistic or economic regions.
\end{abstract}

\ccsdesc[300]{Security and privacy~Human and societal aspects of security and privacy}
\ccsdesc[300]{Information systems~Web applications}


\keywords{ad blocking, filter lists, crowdsource}

\maketitle

\section{Introduction}
\label{sec:intro}

Hundreds of millions of web users (\ie~30\% of all internet users~\cite{adblockUsers}) use filter lists 
to maintain a secure, private, performant, and appealing 
web.  Prior work has shown that filter lists, and the types of content
blocking they enable, significantly reduce data use~\cite{adblock-efficiency},
protect users from malware~\cite{DBLP:conf/ccs/LiZXYW12}, improve browser
performance~\cite{DBLP:conf/websci/GarimellaKM17,Pujol:2015:AUA:2815675.2815705} and significantly reduce how often and persistently
users are tracked on the web.

Most filter lists are generated through crowd-sourcing, where a large
number of people collaborate to identify undesirable online resources
(\eg ads, user tracking libraries, anti-adblocking scripts \etc.) and generate sets of rules to identify those
resources.  Crowd-sourcing the generation of these lists has proven a
useful strategy, as evidenced by the fact that the most popular lists are
quite frequently used and frequently updated~\cite{DBLP:journals/corr/abs-1810-09160,Iqbal:2017:AWR:3131365.3131387}.

The most popular filter lists (\eg EasyList, EasyPrivacy) target ``global''
sites, which in practice means either websites in English, or resources
popular enough to appear on English-speaking sites \emph{in addition to}
sites targeting speakers of other languages.  
Non-English speaking web users face different, generally less appealing options
for content blocking.  Web users who visit non-English websites that target
relatively wealthy users generally have access to well
maintained, language-specific lists.  Indeed, the French~\cite{french-filterlist},
German~\cite{german-filterlist}, and Japanese~\cite{japanese-filterlist} specific filter lists are
representative examples of well-maintained, popular filter lists targeting
non-English web users.
Similarly, linguistic regions with very large numbers of speakers also generally
have well maintained filter lists. Examples here include well maintained
filter lists targeting Hindi~\cite{hindi-filterlist}, Russian~\cite{russian-filterlist}, Chinese~\cite{chinese-filterlist},
and Portuguese~\cite{portugese-filterlist,brazilian-filterlist} websites.

Sadly, users who visit websites in languages with fewer speakers, or with
less wealthy users, have worse options. Put differently, the usefulness of
crowd-sourced filter lists depends on having a large or affluent crowd;
filter lists targeting parts of the web with less, or less affluent, users
are left with filter lists that are smaller, less well-maintained, or
both. Visitors speaking these less-commonly-spoken languages have degraded
web experiences, and are exposed to all the web maladies that filter lists are
designed to fix. 

Compounding the problem, in many cases, users in these
regions are the ones who could benefit most from robust filter lists, as 
network connections may be slower, data may be more expensive, the frequency of undesirable web resources may be higher. An example which motivates this work and illustrates 
the inability of current filter lists to adequately block ads on a regional 
website in Albania can be seen in the screenshot in Figure~\ref{fig:example}. In this example, we browse the website {\it gazetatema.net} while using AdBlock Plus (which uses EasyList, a ``global'' targeting filter list).

While there has been significant prior work on automating the generation of
filter lists~\cite{DBLP:journals/corr/abs-1805-09155,gugelmann2015automated,bhagavatula2014leveraging,7905341}, this existing work is focused on replicating and extending the
most popular English and globally-focused filter lists, with little
to no evaluation on, or applicability to, non-English web regions. 
%
%
In this paper, we target the problem of improving filter 
lists for web users in regions with small numbers of speakers (relative to 
prominent global languages). We select three regions as representative of 
the problem in general: Albania, Hungary and Sri Lanka, using a methodology 
presented in Section~\ref{sec:eval:regions}.

We describe a two-pronged strategy for identifying long-tail resources on
websites that target under-served linguistic regions on the web: (i) a classifier
that can identify advertisements in a way that generalizes well across
languages, and (ii) a method for accurately determining how advertisements
end up in pages (as determined by either existing filter lists or our
classifier), and by using this information, generate new, generalized filter
rules. 

We use this novel instrumentation to both build \emph{inclusions chains}
(\ie measurements of how every remote resource wound up in a web page), and determine
how high in each inclusion chain blocking can begin. This allows us to (i) generate
generalized filter rules (\ie rules that target scripts that include ad images
on each page, instead of rules that target URLs for individual advertisements),
and to (ii) ensure we do not block new resources that will break the website in
other ways.

\point{Contributions}
In summary, this paper makes the following contributions to the problem
of blocking unwanted resources on websites targeting audiences with smaller
linguistic audiences.

\begin{enumerate}[leftmargin=*]
  \item The implementation and evaluation of an \textbf{image classifier for
    automatically detecting advertisements on the web} which relies on a mix of perceptual
    and contextual features.  This classifier is designed to be robust across
    many languages (and particularly those overlooked by existing
    research) and achieves accuracy of \Accuracy{} in identifying images and
    iframes related to advertising. 

  \item Novel, open source \textbf{browser instrumentation, implemented as 
    modifications to the Blink and V8 run-times} that allows for determining
    the cause of every web request in a page, in a way that is far more
    accurate than existing tools.  This instrumentation also allows us to
    accurately attribute every DOM modification to its cause, which in turn
    allows us to predict whether blocking a resource would break a page.
 
  \item The design of a \textbf{novel, two stage pipeline for identifying
    advertising resources on websites}, using the previously mentioned classifier
    and instrumentation, to identify long-tail advertising resources targeting
    web users who do not speak languages with large global communities.

  \item A \textbf{real world evaluation} of our pipeline on sites that
    are popular with languages that are (relatively) uncommon online.
    We find that our approach is successful in significantly improving the
    quality of filter lists for web users without large, language-specific
    crowd sourced lists.
    As our evaluation shows, our generated lists
    block an additional~\NumTotalNewResources{} of ad and ad-related resources (\NumTotalUniqueResources{} unique) when applied to~\NumPagesCrawledAllTotal
    pages targeting these three regions.
\end{enumerate}





\section{Solution Requirements}
\label{sec:motivation:problem}

A successful contribution to the problem of improving the quality
of filter lists in small web regions should account for the following issues:

\point{Scalability}
The primary difficulty of generating effective blocking rules for small-region
web users is the reduced number of people who can participate in a crowd-sourced
list generation.  While portions of the web are targeted at large audiences
(\eg sites in English language, or web regions with a large number
of language speakers) can count on a large number of users to report 
unwanted resources, or generally distribute the task of list generation,
regions of the web targeting only a small number of users (\eg languages
with less speakers) do not have this luxury.  A successful solution therefore
likely requires some kind of automation to augment the efforts of regional
list generators.

\begin{figure}[tb]
    \centering
	\includegraphics[width=0.6\linewidth]{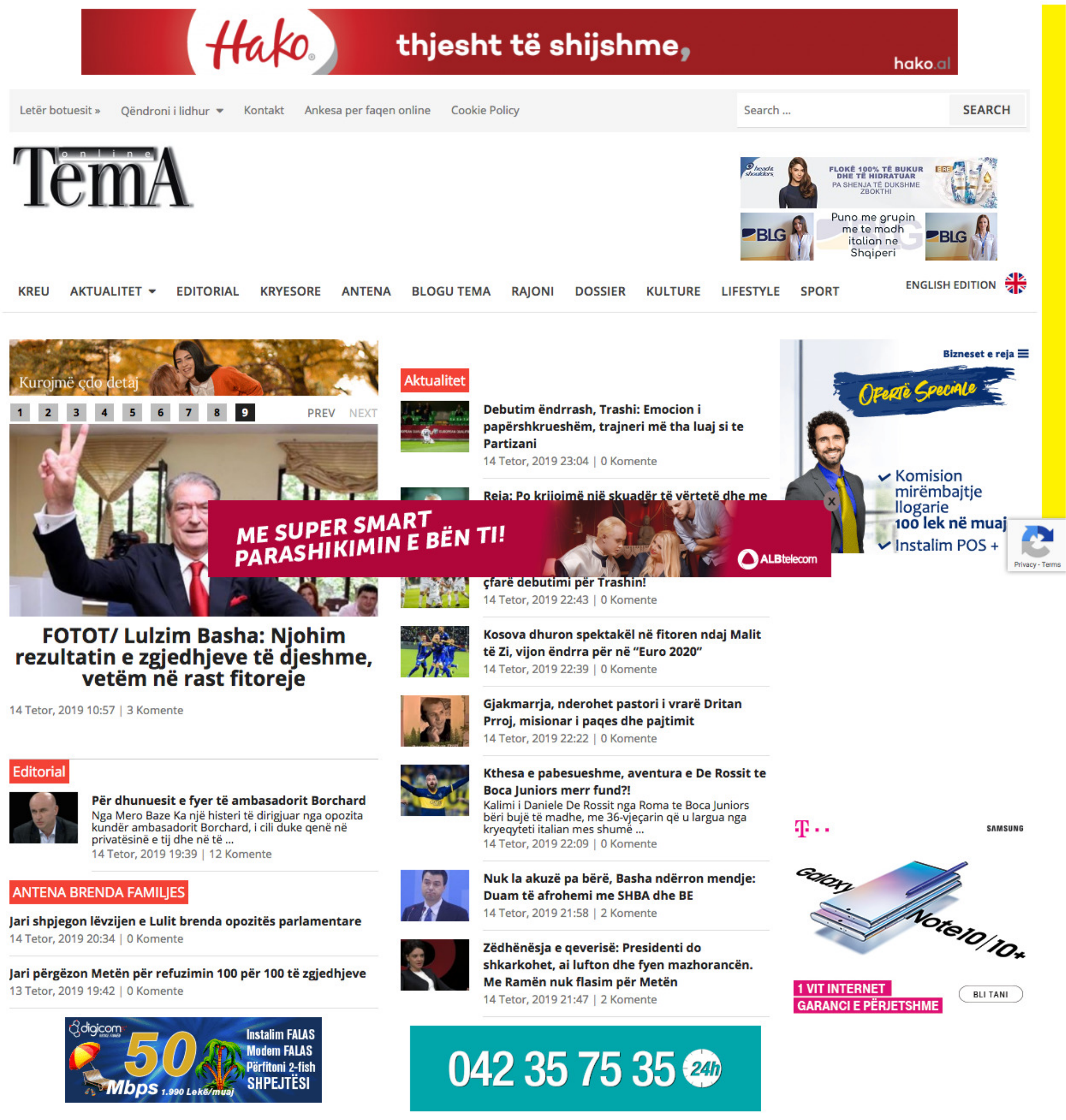}
	\caption{Motivating example of current filter lists' regional inefficiency. Screenshot of Albanian website browsed with Adblock Plus.}
	\vspace{-0.5cm}
	\label{fig:example}
\end{figure}

\point {Generalize-ability}
In most of the cases, ads are rendered by scripts. In addition, every time an 
ad-slot is filled, the embedded ad image may have come from a different 
URL. Approaches that directly target the URLs serving ad-related resources 
are then likely to become stale very quickly. An effective solution to the problem would instead target the ``root cause'' of the unwanted resources being included in the page, in this case the script, which determines what image URLs to load.  Approaches that attempt to only build lists of URLs of ad-related images are therefore unlikely to be useful solutions to the problem for the long term (as seen also from the screenshot in Figure~\ref{fig:example}).

\point{Web compatibility}
Content blocking necessarily requires modifying the execution of a page from
what the site-author intended, to something hopefully more closely aligned
with the visitor's goals and preferences. Modifying the page's execution in this
way (\eg by changing what resources to load, by preventing scripts from executing, \etc)
frequently cause pages to break, and users to abandon content blocking tools.
While filter lists targeting large audiences can rely on the
crowd to report breaking sites to the list authors, (so that they
can tailor the rules accordingly), filter lists targeting smaller
parts of the web often do not have enough users to maintain this positive
feedback loop. An effective system for programmatically augmenting small-region
filter lists must therefore take extra care to ensure that new rules will
not break sites.

\section{Methodology}
\label{sec:meth}

This section presents a methodology for programmatically identifying
advertising and other unwanted web resources in under-served regions.
This section proceeds by describing (i) a high-level overview of our approach,
(ii) a hybrid classifier used to identify image-based web advertisements,
(iii) unique browser instrumentation used in our approach,
(iv) how we identify ad-libraries and other ``upstream'' resources for blocking,
(v) how we determined if a request was safe to block (\ie would not break
desirable functionality on the page), and
(vi) how we generated filter list rules from the gathered ad URLs.

\subsection{Overview}
\label{sec:meth:overview}

Our solution to improving filter lists for under-served regions consists of the combination of two unique strategies. First, we designed a system for
programmatically determining whether an image is an ad, in a cross-language,
highly precise way. We use this classifier to identify ad images that are missed by crowd-sourced filter list maintainers.

Second we developed a technique for identifying additional
resources that should be blocked, by considering the request chains that
brought the ad into the page, and finding instances where we
can block earlier in that request chain. We then apply this ``blocking
earlier in the chain'' principle to both ads identified by existing filter lists, and new ads identified by our classifier, to maximize the number of resources that can be safely blocked.  This approach also allows us to generate generalized blocking rules that target the \emph{causes} of ads being included in the page, instead of only the ``symptoms'': the specific, frequently-changing image URLs.

We note that this approach could be applied to any region of the web, including both
popular and under-served regions. However, since popular parts of the web
are already well-served by crowd-sourced approaches, we expect the marginal
improvement of applying this technique will be greatest for under-served
regions, where there are comparatively few manual labelers.

The following subsections describe the implementation of each piece in our
filter list generation pipeline. Section~\ref{sec:eval} describes the
evaluation of how successful this approach was at generating new filter list rules for under-served regions.

\subsection{Hybrid Perceptual/Contextual Classifier}
\label{sec:meth:classification}

First, our approach requires an oracle for determining if a page element is
an advertisement, without human labeling.  To solve this problem, we designed and trained a unique hybrid image classifier that considers both the image's pixel data, and page context an image request occurred in, when predicting if a page element is an advertisement. Our classifier targets both images (\ie \texttt{<img>})
and sub-documents (\ie \texttt{<iframe>}).  Our classifier prefers precision over recall,
since for filter list it is more important to \textit{only} block ads, instead of blocking \textit{every} ad.

\subsubsection{Comparison to Existing Approaches}
While there is significant existing work on image based (\ie perceptual) web ad classification,
we were not able to use existing approaches for two reasons.  First, we had disappointing results when applying existing perceptual classifiers to the web at large.  The existing approaches we considered did very well on the data sets they
were trained on, but did a relatively poor job when applied to new, random crawls of the
web.

Second, we were concerned that relying on perceptual features alone would
reduce the classifier's ability to generalize across languages.  We expected that
adding contextual features (\eg the surrounding elements in the page, whether the image request was triggered by \JS or the document parser, attributes on the element
displaying the image) would make the classifier generalize better.

\subsubsection{Classifier Design}
\label{sec:meth:classification:design}

Our approach combines both perceptual and contextual page features, each building
on existing work. The perceptual features are similar to those described in the
Percival~\cite{DBLP:journals/corr/abs-1905-07444} paper, while the contextual features
are extensions of those used in the AdGraph~\cite{DBLP:journals/corr/abs-1805-09155} project.
The probability estimated by the perceptual module is then used as an input to the contextual classifier.

\point{Perceptual Sub-module}
The perceptual part of our classifier expands Percival's SqueezeNet based CNN into a larger
network, ResNet18~\cite{he2016deep}.  While the Percival project used a smaller network for fast
online, in-browser classification, our classifier is designed for offline classification, and
so faces no such constraint.  We instead use the larger ResNet18 approach to increase predictive power.
Otherwise, our approach is the same as that described in~\cite{DBLP:journals/corr/abs-1905-07444}.

\point{Contextual Sub-Module}
The contextual part of our classifier does not consider the image's pixel data, but instead 
how the image loaded in the web page, and the context of the page the image or subdocument would be
displayed in. Examples of contextual features include whether the resource being requested is
served on the same domain as the requesting website, and the number of sibling DOM nodes of the
\texttt{img} or \texttt{iframe} element initiating the request. These features are similar to those
described in the AdGraph paper, and detailed in \Cref{fig:classifier-features}.
The browser instrumentation needed to extract these features is described in detailed
in Section~\ref{sec:meth:inst}.

\subsubsection{Classifier Evaluation}
\begin{figure*}[tb]
  \begin{tabular}{lrrrrrr}
    \toprule
      & \multicolumn{3}{c}{Initial Alexa 10k Data Set} & \multicolumn{3}{c}{Alexa 10k Recrawl} \\
    \midrule
      & Accuracy & Precision & Recall & Accuracy & Precision & Recall \\
    \midrule
      Perceptual-only & \perceptualCVAccuracy{} & \perceptualCVPrecision{} & \perceptualCVRecall{} & \perceptualOtherCrawlAccuracy{} & \perceptualOtherCrawlPrecision{} & \perceptualOtherCrawlRecall{} \\
      Hybrid & - & - & - & \hybridCVAccuracy{} & \hybridCVPrecision{} & \hybridCVRecall{} \\
    \bottomrule
  \end{tabular}
  \caption{Comparison of classification strategies. ``Perceptual-only'' refers to the approach by Percival~\cite{DBLP:journals/corr/abs-1905-07444} and variants (best numbers reported).
  ``Hybrid'' uses both perceptual and contextual features, and performed much better on our independent sampling of images and frames from the Alexa 10k, especially with regards to precision.}
	\label{fig:classifier-results}
\end{figure*}

\begin{figure}[tb]
  \begin{tabular}{lrrrr}
    \toprule
      & \# Images & \% Ads & \# Frames & \% Ads \\
    \midrule
      Initial Alexa 10k Set      & \crawlOneTotalImages{} & \crawlOneImagesAdsRatio{} & \crawlOneTotalFrames{} & \crawlOneFramesAdsRatio{} \\
      Alexa 10k Recrawl & \crawlTwoTotalImages{} & \crawlTwoImagesAdsRatio{} & \crawlTwoTotalFrames{} & \crawlTwoFramesAdsRatio{} \\
    \bottomrule
  \end{tabular}
  \caption{Comparison of the distribution of ads for images and frames collected in each data set.}
  \label{fig:classifier-datasets}
\end{figure}
\begin{figure}[tb]
  \begin{tabular}{l}
    \toprule
      Content features \\
    \midrule
      Height \& Width \\
      Is image size a standard ad size? \\
      Resource URL length \\
      Is resource from subdomain? \\
      Is resource from third party? \\
      Presence of a semi-colon in query string? \\
      Resource type (image or iframe) \\
      Perceptual classifier ad probability \\
    \midrule
      Structural features \\
    \midrule
      Resource load time from start \\
      Degree of the resource node (in, out, in+out) \\
      Is the resource modified by script? \\
      Parent node degree (in, out, in+out) \\
      Is parent node modified by script? \\
      Average degree connectivity \\
    \bottomrule
\end{tabular}
  \caption{Partial feature set of the contextual classifier.}
  \vspace{-4mm}
\label{fig:classifier-features}
\end{figure}

We built our image classifier in two steps.  First we built a purely perceptual classifier, using approaches described in existing work.  Second, when we found the perceptual classifier did not generalize well when applied to a new, independent sampling of images, we moved to a hybrid approach.  In this hybrid approach, the output of the perceptual classifier is just one feature among many other contextual features.  We found this hybrid approach performed much better on our new, manually labeled, random crawl of the Alexa 10k.  The rest of this subsection describes each stage in this process. 

Initially, we built a classifier using an approach nearly identical to the perceptual approach described in \cite{DBLP:journals/corr/abs-1905-07444}.  We evaluated this model on a combination of data provided by the paper's authors, augmented with a small amount of additional data labeled by ourselves.  This data set is referred to in Figure~\ref{fig:classifier-datasets} as the ``Initial Alexa 10k Set''.  When we applied the training method described in \cite{DBLP:journals/corr/abs-1905-07444} to this data set, we received very accurate results, reported in Figure~\ref{fig:classifier-results}.

Later, while building the pipeline described in this paper, we generated a second manually labeled data set of images and frames, randomly sampled a new crawl of the Alexa 10k.  This data set is referred to in Figure~\ref{fig:classifier-datasets} as ``Alexa 10k Recrawl'', and was collected between 2-6 months later than the previous data set\footnote{the date range here is due to the majority of this data set being collected by the Percival authors, 6 months before our work, with a smaller additional amount of data being collected by ourselves later on.}.  When we applied the prior purely-perceptual approach to this new data set, we received greatly reduced accuracy.  Most alarming of which, for our purposes, was the dramatically reduced precision.  These numbers are also reported in Figure~\ref{fig:classifier-results}.

We concluded that perceptual features alone were insufficient to handle the breath of advertisements found on the web, and so wanted to augment the prior perceptual approach with additional, contextual features we expected to generalize better, both across languages and across time.  A subset of these contextual features are presented in Figure~\ref{fig:classifier-features}, and are heavily based on the contextual ad-identification features discussed in the AdGraph~\cite{DBLP:journals/corr/abs-1805-09155} project.

After constructing our hybrid classifier from the combination of perceptual and contextual features, we achieved greatly increased precision, though at the expense of some recall. We used a Random Forest approach to combine the perceptual and contextual features, and after conducting a 5-fold cross-validation, achieved mean precision of \hybridCVPrecision{} and mean recall of \hybridCVRecall{}, again summarized in Figure~\ref{fig:classifier-results}.
Our hybrid classifier could not be evaluated against the initial Alexa 10k data set because the data set 1) programmatically determined some labels, and 2) was collected without our browser instrumentation, meaning we could not extract the required features.

\subsection{Browser Instrumentation}
\label{sec:meth:inst}

In this subsection we present \pg, a system for representing and recording web page execution as a graph.  \pg allows us to correctly attribute
every page modification and network request to its cause in the page (usually, the
responsible \JS unit).  We use this instrumentation both to extract the contextual features
described in Section~\ref{sec:meth:classification:design}, and to
accurately understand what page modifications and downstream requests
each \JS unit is responsible for.

Our approach is similar to the AdGraph~\cite{DBLP:journals/corr/abs-1805-09155} project,
but is more robust (\ie corrects categories of attribution errors) and broader
(\ie cover an even greater set of page behaviors).  \pg is implemented as a large
set of patches and modifications to Blink and V8 (approximately~12K LOC).
The code for \pg is open source and actively maintained, and can be found
at~\cite{page-graph},
along with information on how other researchers can use the tool.

The remainder of this subsection provides a high-level summary of the
graph-based approach used by \pg, and how it differs from existing work.

\subsubsection{Graph Representation of Page Execution}
We use \pg to represent the execution of each page as a directed graph.  This
graph is available both at run-time, and offline (serialized as
graphml~\cite{graphml}) for after-the-fact
analysis. \pg uses nodes to represent elements in a web page
(\eg DOM elements, resources requested, executing \JS units, child frames) and edges
representing the interaction between these elements in the page (\eg an edge from
a script to a node might depict the script modifying an attribute on the node, an
edge from a DOM element node to a resource node might depict a file being fetched
because of a \texttt{img} element's \texttt{src} attribute, etc.).  All such page
behaviors in the top-level frame, and child-local-frames, are captured in the graph.

We use \pg's context-rich recording of page execution for several purposes in this work.
First, it allows to accurately and efficiently understand how a \JS unit's execution
modified the page; we can easily determine which scripts made a lot of modifications to
the page, and which had only ``invisible'' effects to \eg fingerprint the user.
Second, the graph allows us to determine how each element ended up in a page.
For example, the graph representation makes it easy to determine if an image was
injected in the page by a script, if so \textit{what} other script, and how
that script was included in the page, etc.  Being able to accurately determine
what page element is responsible for the inclusion of each script, frame or image element
is particularly valuable to this work, as described in the following subsections.

\subsubsection{Differences from Existing Work}

The most relevant related work to \pg is the AdGraph project, which also modifies
the Blink and V8 systems in Chromium to build a graph-representation of page
execution.  \pg differs from AdGraph in several significant ways.

\point{Improved Attribution Accuracy} \pg significantly improves cause-attribution
in the graph, or correctly determining which \JS unit is responsible for each
modification. We observed a non-trivial number of corner cases where AdGraph would attribute
modifications to the wrong script unit, such as when the script was executed as a result
of an element attribute (\eg \texttt{onerror="do\_something()"}), or when the \JS
stack is reset through events like timer callbacks (\eg \texttt{setTimeout(do\_something,1)}).
\pg correctly handles these and a large number of similar corner cases.

\point{Increased Attribution Breadth} \pg significantly increases the set of page
events tracked in the graph, beyond what AdGraph records.  For example, \pg tracks
image requests initiated because of CSS rules and prefetch instructions, records modifications
made in local sub-documents, and tracks failed network requests, among many others. This additional
attribution allows for greater understanding of the context scripts execute
in.

\subsection{Generalizing Filter Rules}
\label{sec:meth:gen-filter-rules}

We next discuss how we generate generalized filter rules from the data gathered
by the previously described image classifier and browser instrumentation. The
general approach is to find URLs serving ad images and frames using the
classifier, use the browser instrumentation to build the entire request
chain that caused the advertisement to be included in the page (\eg
the script that fetched the script that inserted the image), and then
again use the browser instrumentation to determine how far up each
request chain we can block without breaking the page.

We build these request chains for both images (and frames) our classifier identifies as
an ad, and for resources identified by network rules in existing filter lists
(\ie EasyList, EasyPrivacy and the most up to date applicable regional list).
The former allows us to generalize the benefits of our image classifier, the
latter allows us to maximize the benefits of existing filter lists.

\subsubsection{Motivation}
Blocking higher in the request chain has several benefits. First, and most
importantly, targeting URLs higher in the request chain yields a more consistent
set of URLs.  While the specific images that an ad library loads will change
frequently, the URL of the ad library itself will rarely change. Approaches that
target the frequently changing image URLs will result in filter list rules that
quickly go stale; rules that target ad library scripts (as one example) are more
likely to be useful over time, and to a wider range of users. Moving higher in
the request chain means we are more likely to programmatically identify ad
libraries \textit{in addition to} frequently changing, one-off image URLs.

Second, blocking higher in request chains reduces the total number of requests,
bringing privacy and performance improvements.  Blocking a single ``upstream''
ad library may prevent the browser from needing to consider several
``downstream'' requests.

\subsubsection{Building Request Chains}
\begin{figure}[tb]
    \centering
	\includegraphics[width=\linewidth]{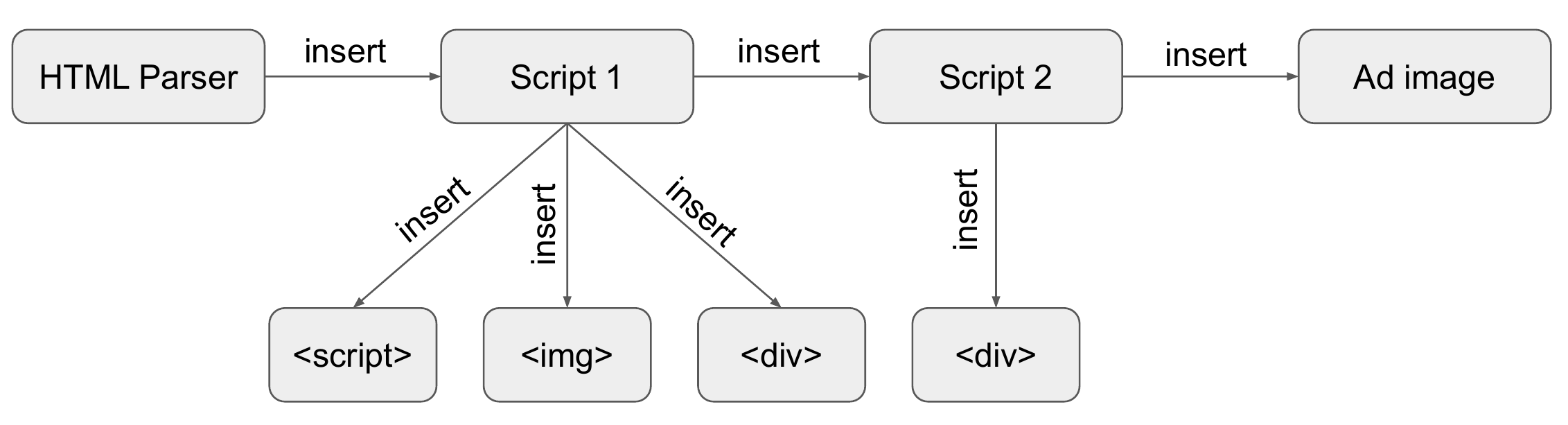}
	\caption{Example of a request chain, ending in an inserted ad image.}
        \vspace{-4mm}
	\label{fig:request-chains}
\end{figure}

To generate optimized filter list rules, we target not only the ad images
and ad frames in each page, but the scripts that injected those images and frames
(and, potentially, the scripts that injected those scripts, etc.). We
refer to the cause of a request as being ``upstream'', and the thing
being requested ``downstream''.  We refer to the list of elements
that participated in an advertisement being included as its ``request chain.''

For each \texttt{<img>}, \texttt{<iframe>} and \texttt{<script>} in a page,
we determine the request chain as follows:

\begin{enumerate}[leftmargin=*]
  \item Locate each element in the \pg generated graph structure.  Call this
    element \texttt{X}.
  \item Use the graph edges to determine how \texttt{X} was inserted in the
    document. If \texttt{X} was inserted by the parser (\ie it appeared in
    the initial HTML text) then stop.
  \item Otherwise, append the script element \texttt{X} into the request
    chain for \texttt{X}, set the responsible script element as the new \texttt{X}
    and continue from \#2 above.
\end{enumerate}

A simplified result of this process is depicted in \Cref{fig:request-chains}.
The figure shows a simplified request chain, where a script was
included in the initial HTML (``script one''), that script programmatically
inserted another script element into the document (``script two'') and that
second script inserted an advertising image into the page.

We use these request chains to determine the optimal place to
start blocking, using the approach described in the following section.

\subsection{Safe Blocking in Request Chains}
\label{sec:meth:safe}
This subsection describes how we determine whether blocking a script
request is likely to break a page.  We use this technique to determine
how ``high'' in each request chain we can block, with the goal of
determining the earliest ``upstream'' request we can block in a request chain
without breaking the page.  Our approach is ``conservative''
(\ie prefers false negatives over false positives), under the intuition that
users would prefer a working, ad-filled page, over a broken, ad-less page.

\subsubsection{Determining Page Breakage}
We use a pair of simple heuristics to determine whether blocking a script
is likely to break a page.  These heuristics are designed to distinguish scripts
that only inject ads into pages from scripts that perform more complex, and hopefully
user serving, page operations.

\begin{enumerate}[leftmargin=*]
    \item If a script creates more than two subtrees in the document, we consider
        it \textbf{unsafe} to block.
    \item If a script inserts another script that matches condition \#1, we consider
        it \textbf{unsafe} to block.
    \item We consider all other scripts \textbf{safe} to block.
\end{enumerate}

Less formally, if a script makes no modifications to the structure of a page,
or the modifications to the page are isolated to one or two parts of the page
(\eg one or two ads, an ad and an ``ad choices'' annotation, etc.) we consider
it safe to block.  Scripts that add elements to more than two parts of the page,
or include scripts that do the same, are considered too risky to block, and too
likely to break desirable page functionality.

We note that this is only a heuristic, one that matches our experience building and debugging
advertising and tracking-blocking tools, but still only a heuristic.  Although heuristics can be fooled by an attacker, they are being used in current tools to identify unwanted code.  If the script for injecting ad content is updated to evade the heuristic, the heuristic can be updated.  We choose the
conservative figure of allowing modifications to a maximum of two regions of the page
to favor false negatives over false positives (\ie we'd rather allow an ad than break a page).
The larger problem of predicting whether any given page modification breaks a site \textit{in the subjective determination
of the browser user} is an open research question, and one that would be its own complicated
project.

\subsubsection{Application to Filter List Generation}
We use the above-described heuristics to determine the highest point in a request
chain that can be blocked.  For each request chain describing how an advertising
image or frame was included in a page, we select the ``highest''
script request we can block, that will not break the page.  Put differently,
we want to select the earliest point in each chain to block, that will have no
``downstream'' breaking scripts.  If there are no elements in the request chain that
can be blocked, the last loaded resource (\ie the ad) will be blocked only.

As a demonstration, consider Figure~\ref{fig:request-chains}.  Our system would
generate two filter rules for this request chain, one targeting the ``ad image'', and one
targeting ``script 2''.
Our system begins by considering the most ``downstream'' request, the image element at the far right.
This image has been identified as
an ad, either by our classifier, or by existing filter lists.  Using the browser
instrumentation described in Section~\ref{sec:meth:inst}, we build the
request chain for this image.

Next, we try to consider the earliest point in the request chain
we can begin blocking.  We observe that ``script 2'' only modifies one other part of the
document (inserting a single \texttt{<div>} element) and so we consider this script
safe to block.  The next element in the request chain, ``script 1'' inserts
elements into more than two parts of the document, and so we consider it
``unsafe'' for blocking.

\subsection{Rule Generation}
\label{sec:meth:rule-generation}

Finally, we describe how we turn the set of identified ad-serving URLs into
filter list rules.  We do so through the following four steps for each URL
we determine to be blockable:

\begin{enumerate}[leftmargin=*]
    \item Reduce the URL's domain to its \texttt{eTLD+1} root.
    \item Remove the query parameter portion of the URL.
    \item Remove the fragment portion of the URL.
    \item Remove the protocol from the URL.
\end{enumerate}
We then record the modified (\ie reduced) version of each URL as a right-rooted AdBlock
Plus format filter rule~\cite{adblock-cheatsheet}.  For example, the URL \texttt{https://a.good.example.com/ad.html?id=3} would
be recorded as \texttt{||example.com/ad.html}, and would block requests to \texttt{https://good.example.com/another-ad.html}
and \texttt{http://a.b.good.example.com/ad.html?id=4}, but would not block
requests to \texttt{https://other.domain.com/ad.html}.  This approach is designed
to generalize some (\ie match other similar requests, even when irrelevant details like tracking
related query parameters change) but not so much so that
unrelated materials served on the same host are blocked.

\section{Evaluation}
\label{sec:eval}

In this section, we evaluate the approach to regional filter
list generation described in \Cref{sec:meth} by applying the technique
to three representative under-served web regions.  We find that the
technique is successful, and generates \NewTotalFilterlistRules
new rules that identify \NumTotalUniqueResources advertising URLs missed by
existing filter lists.
These new rules, when applied in addition to
existing filter list rules, results in \BlockedMoreUniquePercent more
advertising resource being blocked than when using existing filter lists alone.
We also find that our technique is useful
in all three measured regions, though to varying degrees.

This section proceeds by first describing how we selected the three
under-served regions used in this evaluation, then presents the
crawling methodology we used to measure popular websites in each
selected region, and follows by presenting the results of applying
our methodology to each of these regions.  The section concludes by
presenting the output of our measurements (\ie the newly generated filter
list rules), so that they can be used by existing content blockers.

\subsection{Selecting Regions for Evaluation}
\label{sec:eval:regions}
\begin{figure}[tb]
  \begin{tabular}{lrrrr}
    \toprule
        Country  & \# Rules & \# Network Rules & Last Update & Source \\
    \midrule
        Albania & \filterlistrulesAlbania & \filterlistNetworkRulesAlbania & \filterlistUpdatedAlbania & \cite{albania-filterlist} \\
        Hungary & \filterlistrulesHungary & \filterlistNetworkRulesHungary & \filterlistUpdatedHungary & \cite{hungary-filterlist} \\
        Sri Lanka & \filterlistrulesSriLanka & \filterlistNetworkRulesSriLanka & \filterlistUpdatedSriLanka & \cite{sri-lanka-filterlist} \\
    \bottomrule
  \end{tabular}
  \caption{Regional crawling data.  The ``Last Update'' column gives the number of months since the last update, relative to October 2019.}
  \label{fig:filter-lists-by-region}
\end{figure}
\begin{figure}[tb]
  \begin{tabular}{lrrrr}
    \toprule
        Country  & Commits & Start (Month-Year) & Average & Source \\
    \midrule
        Albania & \filterlistCommitsAL & \filterlistStartMonthAL & \filterlistCommitsPerMonthAL & \cite{albania-filterlist} \\
        Hungary & \filterlistCommitsHU & \filterlistStartMonthHU & \filterlistCommitsPerMonthHU & \cite{hungary-filterlist} \\
        Sri Lanka & \filterlistCommitsLK & \filterlistStartMonthLK & \filterlistCommitsPerMonthLK & \cite{sri-lanka-filterlist} \\
        India & \filterlistCommitsHI & \filterlistStartMonthHI & \filterlistCommitsPerMonthHI & \cite{hindi-filterlist-github} \\
        Germany & \filterlistCommitsGE & \filterlistStartMonthGE & \filterlistCommitsPerMonthGE & \cite{german-filterlist-github} \\
        Japan & \filterlistCommitsJA & \filterlistStartMonthJA & \filterlistCommitsPerMonthJA & \cite{japanese-filterlist-github} \\
    \bottomrule
  \end{tabular}
  \caption{Filter list activity.  ``Average''  gives the average number of commits per month since the start of the git repo, relative to Jan. 2020.}
  \vspace{-3mm}
  \label{fig:filter-lists-commits}
\end{figure}

We evaluated our approach on three regions under-served by existing
crowd-sourced filter lists: Albania, Hungary, and
Sri Lanka.  We selected these regions after looking for regions that
matched four criteria.

\begin{enumerate}[leftmargin=*]
    \item The national language was not a major world language.
    \item The amount of updates to the regional supplementary list is significantly lower compared to more popular lists.
    \item The region has seen a vast increase of Internet usage in the past decade\footnote{Statistics gathered from \url{www.internetlivestats.com}}.
    \item Had a popular sites listing on Alexa Top Sites.
    \item There existed at least one filter list for the region.
    \item We could purchase or gain access to a VPN with an exit
        point in the country.
\end{enumerate}

Figure~\ref{fig:filter-lists-by-region} presents the regions we
selected for this evaluation, along with measurements of the existing
best-maintained regional filter list. For each region we identified the best maintained
filter list for the region by consulting both the EasyList selection of regional filter
lists~\cite{adblock-supplement},
and the filter lists indexed on a popular, crowd-sourced site of regional filter lists~\cite{filterlists}.  To further illustrate the lack of updates, \Cref{fig:filter-lists-commits} illustrates how much activity there is on average each month in the respective GitHub repos between the selected regions and some popular filter lists.

\subsection{Crawl Data Set}
\label{sec:eval:data}
\begin{figure}[tb]
  \begin{tabular}{lrrrr}
    \toprule
        Country & \# Domains & \# Pages & \# Images & \# Frames \\
    \midrule
        Albania & \NumDomainsCrawledALTotal & \NumPagesCrawledALUnique & \AllImagesALUnique & \NumFramesObservedUniqueAL \\
        Hungary & \NumDomainsCrawledHUTotal & \NumPagesCrawledHUUnique & \AllImagesHUUnique & \NumFramesObservedUniqueHU \\
        Sri Lanka & \NumDomainsCrawledLKTotal & \NumPagesCrawledLKUnique & \AllImagesLKUnique & \NumFramesObservedUniqueLK \\
    \midrule
        Total & \NumDomainsCrawledAllTotal & \NumPagesCrawledAllUnique & \AllImagesAllUnique & \NumFramesObservedUniqueCombined \\
    \bottomrule
  \end{tabular}
  \caption{Measurements of data gathered from crawls of popular sites in selected under-served regions.  Given numbers are counts of unique image and frame URLs.}
  \vspace{-3mm}
  \label{fig:regional-data-sets}
\end{figure}

We next built a data set of popular websites and pages for each of the three
selected regions.  We use this data set for two purposes: first to approximate how internet
users in these regions experience the web, and second to determine how much
advertising content was being missed by existing content blocking options.

For each region, we first fetched the 1,000 most popular domains for the region,
as determined by the Alexa Top Lists.  Next, we purchased VPN access from
ExpressVPN~\cite{expressvpn}, a commercial VPN service, that provided an IP address in each region.  Third,
we configured a crawler to visit each domain and select two random child links
with the same \texttt{eTLD+1}.  We then configured our crawler to use our
\pg-instrumented browser to visit the domain of each site and each selected
child page, each for 30 seconds.  All crawling was conducted from the VPN
end point, to as closely as possible approximate how the page would
run for a local visitor.

After 30 seconds, we recorded the \pg data for each page (note, the
\pg data includes information about all network requests issued during
the page's execution, in addition to the cause of each request). We also
record all images and scripts fetched in the top-and-local frames
(\ie \texttt{<iframe>}s with the same domain as the top level frame) during each page's execution,
along with screenshots of each remote child-frame (\ie \texttt{<iframe>}s of
third-party domains).

Figure~\ref{fig:regional-data-sets} presents the results of our automated crawl.
For each of the regions we encountered a significant number of non-responsive domains,
which comprise the difference between the number in the ``\# Domains'' column and the
1,000 domains identified by Alexa.  While the 10\% non-responsive rate for Hungary and
Sri Lanka is inline with prior web-studies~\cite{invernizzi2016cloak} that find around ~11\% error rate for
automated crawls of the web, the even higher rate of non-responsive sites in Albania
was surprising.  On manual evaluation of a sample of these domains, we found
a small number of cases were due to anti-crawler countermeasures or apparent IP
blacklisting of the VPN end point. In a surprising number of cases though, domains
seemed to be abandoned and hosting no web content at all.  We note this as a point
for future study.

\subsection{Ads Identified by Existing Filter Lists}
\label{sec:eval:list-ads}
\begin{figure}[tb]
  \begin{tabular}{lrrrr}
    \toprule
      Country & \# Ad Images & \% & \# Ad Frames & \% \\
    \midrule
      Albania & \ImagesDirectFromListsALUnique & \ImagesDirectFromListsALUniquePct & \FramesDirectFromListsALUnique & \FramesDirectFromListsALUniquePct \\
      Hungary & \ImagesDirectFromListsHUUnique & \ImagesDirectFromListsHUUniquePct & \FramesDirectFromListsHUUnique & \FramesDirectFromListsHUUniquePct \\
      Sri Lanka & \ImagesDirectFromListsLKUnique & \ImagesDirectFromListsLKUniquePct & \FramesDirectFromListsLKUnique & \FramesDirectFromListsLKUniquePct \\
    \midrule
      Total & \ImagesDirectFromListsCombinedUnique & \ImagesDirectFromListsCombinedUniquePct & \FramesDirectFromListsCombinedUnique & \FramesDirectFromListsCombinedUniquePct \\
    \bottomrule
  \end{tabular}
  \caption{Measurements of how many unique image and iframe ads are currently identified by existing filter lists (\eg EasyList, EasyPrivacy, and the best maintained filter list for each region.).}
  \label{fig:regional-ads-filter-lists}
\end{figure}

Next, we measured how successful existing filter lists are at blocking
advertisements on popular sites in our selected regions.  We treated this
measurement as our baseline when measuring how much additional blocking benefit our
approach provides.  Figure~\ref{fig:regional-ads-filter-lists} shows the number of ads
identified by existing filter lists.

We measured the amount of ad resources identified by existing lists in two steps.
First, we combined the best maintained regional filter list for each region (listed
in Figure~\ref{fig:filter-lists-by-region}) with the two most popular ``global'' filter
lists, EasyList and EasyPrivacy.  Then, we applied these combined filter lists to
the image and iframe requests encountered when crawling each region, using a popular AdBlock
filter list library~\cite{adblock-rust}.  We then noted which images and iframe requests
would be blocked by the current best filter lists available to people in each region.

\subsection{Ads Identified by Hybrid Classifier}
\label{sec:eval:classifier-ads}
\begin{figure}[tb]
  \begin{tabular}{lrrrr}
    \toprule
      Country & \# Ad Images & \% & \# Ad Frames & \% \\
    \midrule
      Albania & \ImagesDirectFromOnlyUsALUnique & \ImagesDirectFromOnlyUsALUniquePct & \FramesDirectFromOnlyUsALUnique & \FramesDirectFromOnlyUsALUniquePct \\
      Hungary & \ImagesDirectFromOnlyUsHUUnique & \ImagesDirectFromOnlyUsHUUniquePct & \FramesDirectFromOnlyUsHUUnique & \FramesDirectFromOnlyUsHUUniquePct \\
      Sri Lanka & \ImagesDirectFromOnlyUsLKUnique & \ImagesDirectFromOnlyUsLKUniquePct & \FramesDirectFromOnlyUsLKUnique & \FramesDirectFromOnlyUsLKUniquePct \\
    \midrule
      Total & \ImagesDirectFromOnlyUsCombinedUnique & \ImagesDirectFromOnlyUsCombinedUniquePct & \FramesDirectFromOnlyUsCombinedUnique & \FramesDirectFromOnlyUsCombinedUniquePct \\
    \bottomrule
  \end{tabular}
  \caption{Measurements of how many unique image and iframe ads the classifier described in Section~\ref{sec:meth:classification} identified that \textit{were not identified} as ads by existing filter lists.}
  \vspace{-5mm}
  \label{fig:regional-ads-classifier}
\end{figure}

Next, we identified how many advertising images and iframes we missed by
existing filter lists.  We found a significant number of both; our classifier
identified \ImagesDirectFromOnlyUsCombinedUnique ad images and \FramesDirectFromOnlyUsCombinedUnique
ad frames that were missed by existing filter lists.  Put differently, our approach identified \ImagesDirectFromOnlyUsCombinedUniquePct more ad images, and \FramesDirectFromOnlyUsCombinedUniquePct more ad frames, than existing filter lists.

We note that these figures only include ad images and frames missed by filter lists;
we observed a significant amount of overlap between the two approaches.  Figure~\ref{fig:regional-ads-classifier}
summarizes the additional ad resources our classifier identified.

We measured how many advertising images and frames current approaches miss in
two steps.  First, we identified each image or frame in the corresponding \pg
graph data, and used that contextual information to extract the contextual features
described in Section~\ref{sec:meth:classification:design}. Second, we used the pixel
data of each resource to feed the perceptual part of the classifier.  For images,
we used the image file directly; for frames we used a screenshot of the frame
taken during the crawling step.

\subsection{Generalizing Rules with Request Chains}
\label{sec:eval:extend}
\begin{figure}[tb]
  \begin{tabular}{lrrrr}
    \toprule
      Country & Current Lists & Classifier & $\cup$ Chains & $\Delta$ \\
    \midrule
      Albania & \CombinedDirectFromListsALUnique & \CombinedDirectFromOnlyUsALUnique & \NumTotalUniqueResourcesAL & \BlockedMoreUniquePercentAL \\
      Hungary & \CombinedDirectFromListsHUUnique & \CombinedDirectFromOnlyUsHUUnique & \NumTotalUniqueResourcesHU & \BlockedMoreUniquePercentHU \\
      Sri Lanka & \CombinedDirectFromListsLKUnique & \CombinedDirectFromOnlyUsLKUnique & \NumTotalUniqueResourcesLK & \BlockedMoreUniquePercentLK \\
    \midrule
      Total & \CombinedDirectFromListsAllUnique & \CombinedDirectFromOnlyUsCombinedUnique & \NumTotalUniqueResources & \BlockedMoreUniquePercent \\
    \bottomrule
  \end{tabular}
  \caption{Additions to filter lists when applying all steps of our methodology.
    ``Current lists'' gives the number of ad-resources found by existing filter lists, ``classifier'' describes ad-resources found by our hybrid classifier but not existing filter lists. ``$\cup$ chains'' gives the number of new ad-resources found by applying our ``upstream'' approach to ad-resources found by either current filter lists or the hybrid classifier.  The ``$\Delta$'' column gives the overall increase in identified ad-resources provided by our techniques, compared to existing filter lists.}
  \vspace{-5mm}
  \label{fig:regional-ads-combined}
\end{figure}

Next, we identified additional resources that should be blocked by examining the
request chain for each ad image or frame, and finding the earliest point in each
chain that could be blocked without breaking the page.  We applied this ``upstream request
chain'' blocking technique to both ad resources labeled by existing filter lists and
ad resources newly identified by our hybrid classifier.  Doing so allowed us to
not only identify specific images and frames that should be blocked, but to programmatically
identify the ``upstream'' libraries that caused those images and frames to be included.

We were able to identify \NumTotalUniqueResources additional advertising URLs by analyzing
the request chains in this manner, an improvement of \BlockedMoreUniquePercent in advertisement
blocking in these regions.  We note that by following the methodology described in
Section~\ref{sec:meth:safe}, targeting these additional resources will result in more generalizable
filter rules by identifying both
the individual ad image URLs \textit{and} the ad libraries that determine what ads to load.
The approach described in Section~\ref{sec:meth:safe} also gives us a high degree of
confidence that this ``upstream'' blocking will not break pages.

Figure~\ref{fig:regional-ads-combined} shows the final results of our regional filter
list methodology when applied to the Albanian, Hungarian and Sri Lankan web regions.
The ``current lists'' column presents the number of ad resources
existing filter lists identify in each region in our data set.  The ``classifier''
column gives the number of images and frames our hybrid classifier identifies as ad-related
\textit{that are not identified} by existing filter lists.  The ``$\cup$ chains''
column gives the total number of ad resources identified by applying
the request chain approach (Section~\ref{sec:meth:gen-filter-rules}) to images
and frames identified as advertisements by \textit{either} existing filter lists or
the hybrid classifier.  The final ``$\Delta$'' column gives the percent-increase
in resources identified by our combined methodology, when compared to using only
existing filter lists.

\subsection{Generated Filter-Lists}
\label{sec:eval:output}
\begin{figure}[tb]
  \begin{tabular}{lr}
    \toprule
    {\bf  Country} & {\bf  Network rules} \\
    \hline
    Albania &  \NewALFilterlistRules \\
    Hungary &  \NewHUFilterlistRules \\
    Sri Lanka & \NewLKFilterlistRules \\
    \hline
    Total & \NewTotalFilterlistRules \\
    \bottomrule
  \end{tabular}
  \caption{Number of new filter list rules.}
  \vspace{-0.5cm}
  \label{fig:regional-new-rules}
\end{figure}

Finally, we generated filter lists in AdBlock Plus format to block the advertising resources
identified in the previous steps.  We used the rule generation methodology described
in Section~\ref{sec:meth:rule-generation} to generate \NewTotalFilterlistRules new filter rules.
We have made the filter lists available~\cite{generated-filter-lists}.  Counts of the total number of new rules for each region are presented in Figure~\ref{fig:regional-new-rules}.

\section{Discussion}
\label{sec:discussion}
In this section, we discuss broader issues related to the problem of filter
list generation and ad blocking, including possible next steps and extensions for the described
approach, and some limitations and concerns for future researchers
to consider.

\subsection{Ad Ambiguity}
\label{sec:discussion:ambiguity}
A reoccurring issue in identifying and blocking online advertisements
is that many images and ads are context specific. An ad in
one context might be core content in another. For example,
an image of shoes with the name of the shoe maker might be
perceived as an advertisement when positioned next to a news
article, but the same image might be desirable when placed in the
middle of a page on a shoes selling website.

\begin{figure}[t]
    \centering
	\includegraphics[width=0.35\linewidth]{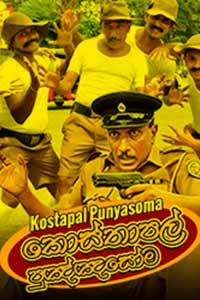}
	\caption{Example of an ambiguous ad image we encountered on a movie sharing forum in Sri Lanka as part of the its table of contents.}
	\vspace{-0.5cm}
	\label{fig:ambigiousAd}
\end{figure}

We encountered an even more difficult case when labeling and debugging
the pipeline described in this work.  We found a movie sharing
forum that used a number of banner ads (like the one presented in 
Figure~\ref{fig:ambigiousAd}) from elsewhere on the web as a table
of contents, to show which movies had most recently been added on 
the site.  In such cases, the ``ad-ness'' of an image
is not ambiguous, its explicitly both an ad and desirable page content!

Crowd-sourced filter list generation approaches rely on the
subjective intuition of list contributors to resolve
such difficult situations.  Programmatic solutions have
no such option, and so must address a two tiered problem: first,
how to identify images that look like advertisements, and second,
how to model subjective user expectations of when an advertisement is
desirable to users.

We find the problem presented by the intersection of these
two issues is unaddressed by existing literature (current work included).
Our resolution to this issue was ``ads are ads''', and it is the
job of ad blocking tools to block ads, and if a user is in a scenario where
they wish to see and ad, they should disable the ad-blocking tool.  How satisfying
such an approach is will likely be task specific.  Thankfully,
the number of ambiguous ads we encountered was low enough that it
did not affect the main focus of the work, but we mention it as an interesting
area for future research.

\subsection{Alternative Image-Identification Oracles}
\label{sec:discussion:alts}

This work used a unique hybrid approach for determining whether an
image or a frame was an advertisement.  This is only one of an infinite number
of possible oracles the same pipeline could adopt.  While we designed our approach
to be conservative in identifying images (as described in Section~\ref{sec:meth:classification}),
one could instead use a much more aggressive oracle, if one was willing to accept
a greater false-positive rate in ad-identification, or was willing to
accept sites breaking, for additional data savings and privacy protections.

In this sense, our oracle represents just one web use preference (less advertisements, but
with a low tolerance for error).  The same broad approach, as described in this paper,
could be used with other oracles, such as those targeting just certain types of
advertisements (\ie blocking adult ads), or certain types of web content in general
(\ie blocking violent images).

While our goal in this work was to improve web browsing for
people in under-served regions, the described approach is not specific to advertising.
The identify-and-prune-the-request-chain approach could be helpful in addressing many
web problems where human labelers are lacking.

\subsection{Possible Extensions}
\label{sec:discussion:improvements}

We considered many additional features and approaches when designing the methodology
in this work.  Here, we briefly describe a variety of improvements we considered,
but did not implement because of time, cost or complexity.  We list them
as possible suggestions to other researchers addressing similar problems.

\point{Predicting Page Breakage} An important part of this work was generating and
testing useful heuristics for whether blocking a script would
break a page.  The heuristics discussed in Section~\ref{sec:meth:safe} have proven useful
for us, but could be improved.  One could, for example, also consider the number and type
of Web API calls a script makes, whether the script sets or reads storage, or any
number of other behavioral characteristics when trying to predict whether blocking
a script would break a page.

\point{Tracking Protections} This work improves blocking \textit{advertisements} in under-served
regions, but similar approach could be taken to target \textit{tracking scripts}.  Instead
of building a classifier to determine if an image is an advertisement, one would instead
need an oracle to determine whether a script was privacy violating.  This might be an easier task,
since determining the privacy implications of a script's execution is in many cases easier
that predicting the subjective evaluation of whether an image is an advertisement.

\point{Improving Other Filter Lists} The approach in this paper was designed to help web users
in under-served regions.  However, the same approach could likely be used to improve
filter lists in general, including ``global'' popular ones like EasyList and EasyPrivacy.
Though the marginal improvement would likely be lower, since the relative popularity of
such sites likely means a higher percentage of ad resources have been identified by
filter list contributors, our approach could still be useful in improving blocking
on less popular, or frequently changing sites.

\subsection{Limitations}
\label{sec:discussion:limits}

Finally, we note some limitations of this approach, in the hopes that future work
might address them.  First, while our approach was successful on the three selected
regions, its difficult to know if these findings would generalize to all under-served
regions on the web.  While such a measurement is beyond our ability to carry out,
it would be interesting to better understand how similar web advertising is across
the web generally.

Second, our approach relies on automated, manual crawls of websites to identify
ad-related resources.  It is possible that the kinds of advertisements reachable by
automated tools are different from the kinds of advertisements humans experience
when on parts of the web not reachable by crawlers, such as within web applications,
behind paywalls, or within account-requiring portions of websites.  This limitation
is a subset of a larger open problem in web measurement, of understanding how well
automated crawls approximate human user experiences.

Finally, the types of advertisements targeted in this work (\ie image based web advertisements)
are just one of many types of advertisements web users face.  A partial list includes
audio ads, video interstitials, native text ads, and interactive advertisements.
If image- and frame- targeting ad blockers continue to become more popular, we can
expect advertisers to adopt to these alternative advertising approaches.  Researchers
will in turn need to come up with new ad blocking techniques to preserve a usable, performant,
privacy respecting web.

\section{Related Work}
\label{sec:back}
Below we cover the existing work related to filter lists (Section~\ref{sec:back:lists}),
resource blocking (Section~\ref{sec:back:blocking}),
and the importance of different vantage points for web measurements 
(Section~\ref{sec:back:crawling}).

\subsection{Filter Lists}
\label{sec:back:lists}

In~\cite{DBLP:journals/corr/abs-1810-09160}, Vastel et al. explored the accumulation of dead rules by studying EasyList, the most popular filter list.
Results of their study show that the list has grown from several hundred rules, to well over~60,000 rules, within 9 years, when~90.16\% of the resource blocking rules are not useful. Finally, authors, propose optimizations for popular ad-blocking tools, that allow EasyList to be applied on performance constrained mobile devices, and improve desktop performance by 62.5\%.

Gugelmann et al. in~\cite{gugelmann2015automated} investigated how to detect privacy-intrusive trackers and services from passive measurements and propose an automated approach that relies on a set of web traffic features to identify such
services and thus help developers maintaining filter lists.
Pujol et al. in~\cite{Pujol:2015:AUA:2815675.2815705} used Adblock Plus filter lists for passive network classification.  By analyzing 
data from a major European ISP authors show that 22\% of the active users have Adblock Plus deployed. Also 
they found that  56\% and 35\% of the ad-related requests are blacklisted by EasyList and EasyPrivacy, respectively.

Iqbal et al., in~\cite{Iqbal:2017:AWR:3131365.3131387}, studied the anti-adblock filter lists that ad blockers use to remove anti-adblock scripts. By analyzing the evolution of two popular anti-adblock filter lists, authors show that their coverage considerably improved the last~3 years and they are able to detect anti-adblockers on about~9\% of Alexa top-5K websites. Finally authors proposed a machine learning based method to automatically detect anti-adblocking scripts.

\vspace{-2mm}
\subsection{Resource Blocking}
\label{sec:back:blocking}
Iqbal et al., in~\cite{DBLP:journals/corr/abs-1805-09155}, proposed AdGraph: a graph-based machine learning approach for detecting 
advertising and tracking resources on the web. Contrary to filter list based approaches AdGraph builds a graph representation of the HTML structure, network requests, and JavaScript behavior of a webpage, and uses this unique representation to train a classifier 
for identifying advertising and tracking resources. AdGraph can replicate the labels of human-generated filter lists with~95.33\% accuracy.

In~\cite{storey2017future}, Storey et al. discussed the future of ad blocking by modelling it as a state space with four states and six state transitions, which correspond to techniques that can be deployed by either publishers or ad blockers. They also proposed several new ad blocking techniques, including ones that borrow ideas from rootkits to prevent detection by anti-ad blocking scripts. 
Zhu et al., in~\cite{Zhu:2019:SLS:3308558.3313558}, proposed ShadowBlock: a new Chromium-based ad-blocking browser that 
can hide traces of ad-blocking activities from anti-ad blockers. ShadowBlock leverages existing filter lists and hides all ad elements 
stealthily so anti-ad blocking scripts cannot detect any tampering of the ads (\eg, absence of ad elements). Performance evaluation on 
Alexa top-1K websites shows that their approach successfully blocks 98.3\% of all visible ads while only causing minor
breakage on less than~0.6\% of the websites.

Garimella et al. in~\cite{DBLP:conf/websci/GarimellaKM17} measured the performance and privacy aspects of popular ad-blocking tools. 
Their findings show that (i) uBlock has the best performance, in terms of ad and third party tracker filtering, and least privacy
tracking. They also found that the time to load pages is not necessarily faster when using adblockers, and this happens due to additional
functionality introduced by the adblocking tools.
In~\cite{DBLP:journals/corr/abs-1905-07444}, Din at al. proposed Percival: a deep learning based perceptual ad blocker that 
aims to replace filter list based adblocking. Percival runs within the browser's image rendering pipeline, intercepts images during page execution and by 
performing image classification, it blocks ads. Percival can replicate 
EasyList rules with an accuracy of~96.76\% when it imposes a rendering performance overhead of~4.55\%.


\subsection{Internet Vantage Points}
\label{sec:back:crawling}
Selecting different vantage points to browse Internet from is a quite common technique in order to understand the different view of 
the web different users may have. Jueckstock et al. in~\cite{DBLP:journals/corr/abs-1905-08767} design and deploy a 
synchronized multi-vantage point web measurement study to explore the comparability of web measurements across 
different Internet vantage points.
In~\cite{DBLP:journals/corr/FruchterMSB15} Fruchter et al. proposed a method for investigating tracking behavior by analyzing 
cookies and HTTP requests from browsing sessions from different countries. Results show that websites track users
differently, and to varying degrees, based on the regulations of the country the visitor's IP is based in. 

Iordanou et al. in~\cite{Iordanou:2017:FPB:3098822.3098850} proposed a system for measuring how e-commerce 
websites discriminate between users. Authors consider several different motivations for discrimination, including 
geography, prior browsing behavior (e.g., tracking-derived PPI) of the user, and site A/B testing. 
They found that the first and third motivations explain more website ``discrimination'' than the second.

\vspace{-2mm}
\section{Conclusions}
\label{sec:conclusion}

In this work we address the problem of augmenting filter lists for users of
under-served, linguistically-small parts of the web.  
The approach described in this work is amenable to full automation, and with sufficient computation resources could be applied to any number of additional under-served populations of web users.  Further, we expect the same approach could be used to block tracking-related resources too, improving privacy for under-served web users too.  

The problem of poorly maintained filter lists in under-served regions is significant. First, the current predominant approach
to filter list generation (\ie crowd-sourcing) is poorly suited for these web-regions, which by definition have less users, and so smaller ``crowds.'' Second, in many cases, under-served areas of the web target users with less income, and with less access to cheap, high speed data; the users who would benefit most from ad blocking are often the poorest served by current filter list generating strategies. Third, existing filter list generation strategies that \emph{do not} rely on crowd-sourcing fail to consider web compatibility (\ie breaking sites), leaving under-served users with the unappealing trade-off between data-draining, privacy-harming browsing, or, alternatively, breaking web sites.

This work proposes a novel approach for generating filter rules for under-served regions of the web.  Our approach determines whether images and frames are advertisements by considering perceptual and contextual aspects of the underlying image (or frame), and then using deep browser instrumentation to determine where in the request chain we can optimally begin blocking requests.

We apply this approach to popular websites in three regions currently poorly served
by crowd-sourced filter lists, Sri Lanka, Hungary, and Albania. Our approach is successful at improving blocking without breaking websites.  We generate \NewTotalFilterlistRules new filter list rules that identify \BlockedMoreUniquePercent new advertising resources that should be blocked, improving blocking by \BlockedMoreTotalPercent over the existing best options for these regions.  We are also releasing our generated filter lists so that web users in these regions can benefit from them~\cite{generated-filter-lists},
along with the source code for our hybrid image classifier~\cite{filterlist-gen-page} and our \pg browser instrumentation~\cite{page-graph}.
We hope this work advances the goal of improving the web for all users, no matter their location or linguistic-community.

\begin{acks}
This work was partly funded by the Swedish Foundation for Strategic Research (SSF) and the Swedish Research Council (VR).
\end{acks}

\bibliographystyle{plain}
\bibliography{paper}
\end{document}